\documentclass[nopacs,twocolumn,preprintnumbers,amsmath,amssymb,superscriptaddress,nofootinbib]{revtex4}\usepackage[dvips]{graphicx}
\usepackage{amssymb}
\usepackage{amsmath}
\usepackage{graphicx,color}

\usepackage{times}

\usepackage{enumerate}
\usepackage{mathtools}
\usepackage{stmaryrd}

\usepackage{epstopdf}

\usepackage{textcomp}
\usepackage{gensymb}

\usepackage{footmisc}

\newcommand{\moy}[1]{\left\langle #1 \right\rangle}

\newcommand{\ex}[1]{\mathrm{e}^{#1}}

\newcommand{\RR}[0]{\boldsymbol{R}}
\newcommand{\CC}[0]{\boldsymbol{\mathcal{C}}}
\newcommand{\xx}[0]{\boldsymbol{x}}

\newcommand{\kB}[0]{k_{\mathrm{B}}}

\newcommand{\uu}[0]{\hat{\boldsymbol{u}}}

\newcommand*\samethanks[1][\value{footnote}]{\footnotemark[#1]}

\begin{document}

\title{Exothermicity is not a necessary condition for enhanced diffusion of enzymes}

\author{Pierre Illien\footnote{P.I. and X.Z. contributed equally to this work.}}
\address{Rudolf Peierls Centre for Theoretical Physics, University of Oxford, Oxford OX1 3NP, UK}
\address{Department of Chemistry, The Pennsylvania State University, University Park, PA 16802, USA}

\author{Xi Zhao\samethanks{}}
\address{Department of Chemistry, The Pennsylvania State University, University Park, PA 16802, USA}

\author{Krishna K. Dey\footnote{Present address: Indian Institute of Technology Gandhinagar, Palaj Campus, Gandhinagar, Gujarat 382 355, India.}}
\address{Department of Chemistry, The Pennsylvania State University, University Park, PA 16802, USA}

\author{Peter J. Butler}
\address{Department of Biomedical Engineering, The Pennsylvania State University, University Park, PA 16802, USA}

\author{Ayusman Sen\footnote{Corresponding authors: asen@psu.edu,\\ ramin.golestanian@physics.ox.ac.uk}}
\address{Department of Chemistry, The Pennsylvania State University, University Park, PA 16802, USA}

\author{Ramin Golestanian\samethanks{}}
\address{Rudolf Peierls Centre for Theoretical Physics, University of Oxford, Oxford OX1 3NP, UK}

\date{\today}

\begin{abstract}

Recent experiments have revealed that the diffusivity of exothermic and fast enzymes is enhanced when they are catalytically active, and different physical mechanisms have been explored and quantified to account for this observation. We perform measurements on the endothermic and relatively slow enzyme aldolase, which also shows substrate-induced enhanced diffusion. We propose a new physical paradigm, which reveals that the diffusion coefficient of a model enzyme hydrodynamically coupled to its environment  increases significantly when undergoing changes in conformational fluctuations in a substrate-dependent manner, and is independent of the overall turnover rate of the underlying enzymatic reaction. Our results show that substrate-induced enhanced diffusion of enzyme molecules can be explained within an equilibrium picture, and that the exothermicity of the catalyzed reaction is not a necessary condition for the observation of this phenomenon.

\end{abstract}

\maketitle

In a quest for understanding nonequilibrium processes encountered in biology and chemistry, the study of active matter, namely systems constituted of agents able to consume and convert energy extracted from their environment, has been a major focus of the contemporary physical sciences \cite{Marchetti2013, Bechinger2016}. Recent progress led  to the design, fabrication and characterization of synthetic micro- and nano-machines relying on different propulsion mechanisms, and able to reproduce functions inspired from molecular biology, such as cargo transport or chemical sensing \cite{Sanchez2015,Illien2017}. Such autonomous objects could have major technological applications, provided that they are small enough and fully biocompatible. In this context, and going down in scale, enzyme molecules have received a lot of attention, as models of biological nanoscale transducers able to convert chemical energy into mechanical work. Biomolecules typically perform cyclic turnovers in which they bind to substrates and catalytically convert them to products while undergoing conformational changes \cite{Grosberg1994,Alberts2002,Phillips2008,Nelson2008}. Recently, in vitro studies of enzymes using fluorescence correlation spectroscopy (FCS) have revealed that their diffusion coefficient is enhanced in a substrate-dependent manner \cite{Muddana2010a,Sengupta2013,{Sengupta2014a},Riedel2015}, and that the diffusion enhancement $\Delta D$ at substrate saturation was typically of the order of the bare diffusion coefficient of the enzyme $D_0$ measured in the absence of substrate molecules. This observation holds for a wide range of enzymes, which typically catalyze fast and exothermic chemical reactions, with reaction enthalpies that can reach 40$\kB T$ per molecule and catalytic rates up to $\sim 10^4$ s$^{-1}$ for the particular case of catalase \cite{Riedel2015}.

This intriguing phenomenon, that could have major implications in the spatial organisation of biological processes \cite{Wu2015}, was subsequently investigated from a theoretical point of view. It was first suggested that the enhancement of the enzymes diffusion coefficient is directly proportional to the overall rate of the catalytic reaction, and that there is a correlation between the degree of exothermicity of the overall reaction and the observed enhancement in diffusion \cite{Riedel2015}. In support of these findings, a theoretical scenario was proposed, in which the energy released by the chemical reaction is assumed to be channeled into an asymmetric compression of the molecule and converted into a translational boost. However, the theoretical picture  proposed in support of these experimental findings was subsequently criticized, as it relies on an underestimate of the friction coefficient of the protein and on the hypothesis that the released energy is partitioned only over a small number of degrees of freedom \cite{Golestanian2015}. Alternatively, we recently proposed that the exothermicity of the reaction catalyzed by the enzymes was responsible for collective heating of the sample container, that could contribute to the enhanced diffusion of the enzyme molecules \cite{Golestanian2015}.

The role played by stochastic swimming of enzyme molecules induced by conformational changes was also investigated within a nonequilibrium picture \cite{Golestanian2008,Sakaue2010,{Golestanian2015},{Bai2015}}. With a simplified description of the mechanochemical cycle of the enzyme, it was shown that the diffusion enhancement was controlled by the overall catalytic rate of the reaction $k_\text{cat}$ through the relation $\Delta D \sim k_\text{cat} R^2$ where $R$ is the hydrodynamic radius of the enzyme, and represents an upper bound for the typical length scale representing the magnitude of its conformational changes \cite{Golestanian2015}. However, even for fast enzymes such as catalase, the relative change in the diffusion coefficient barely reaches the orders of magnitude observed in experiments \cite{Golestanian2015}. It was finally proposed that enzymes could act as active force dipoles, that create non-thermal fluctuating solvent flows, and that could be responsible for enhanced diffusion \cite{Mikhailov2015}. In such a collective picture, the diffusion change is controlled by the volume fraction of enzymes in the sample, which are usually very small in the FCS experiments. Consequently, although such effects could potentially have important consequences for denser suspensions, they cannot account for the experimental realisations mentioned above.

Therefore, the status quo of the physical understanding of this phenomenon is that it is an intrinsically nonequilibrium process, and relatively satisfactory explanations were only proposed for enzymes which are  sufficiently fast or  catalyze sufficiently exothermic reactions. In search of a more complete physical picture, it is pertinent to probe whether exothermicity is a necessary condition for the phenomenon, and whether the enhanced diffusion is controlled by the overall catalytic rate. To this end, we experimentally studied aldolase, an enzyme involved in different fundamental metabolic processes such as glycolysis, since it has the following properties: First, this enzyme is known to be endothermic, with a reaction enthalpy estimated ranging from 30 to 60 kJ/mol \cite{Minakami1976,Goldberg1995}. Secondly, the turnover rate of this enzyme is very low, with a maximum of 5 product molecules generated per second at substrate saturation \cite{Callens1991}. Aldolase converts its substrate fructose-1,6-bisphosphate (FBP) into the products dihydroxyacetone-phosphate (DHAP) and glyceraldehyde 3- phosphate (G3P).

\begin{figure}
\begin{center}
\includegraphics[width=\columnwidth]{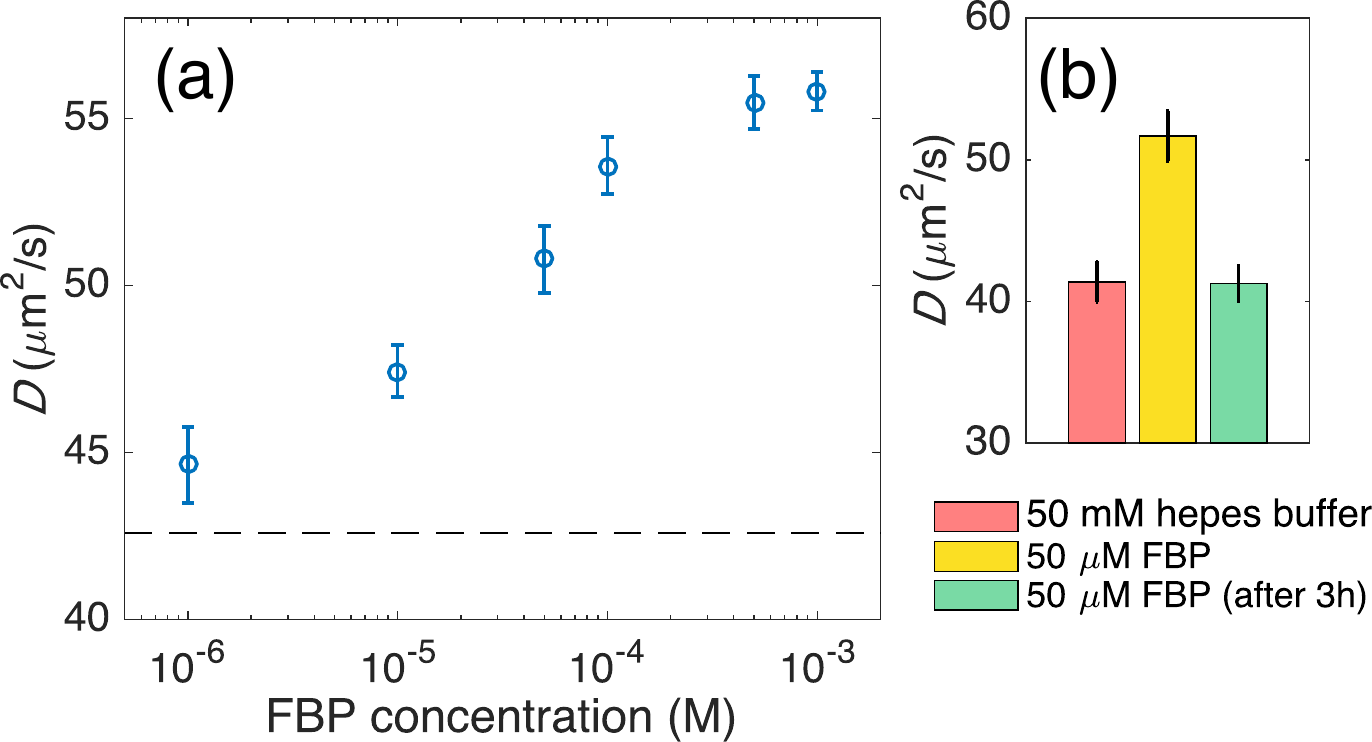}
\caption{(a) Diffusion coefficient of aldolase molecules measured in experiments (the error bars represent standard deviations calculated for 15 different measurements under identical conditions) as a function of  FBP (substrate) concentration. ({b}) The enhanced diffusion of aldolase in the presence of substrate returns to the base value (observed in the absence of the substrate) when the substrate is consumed. All values are significantly different with $p <0.05$.}
\label{fig_exp_substrate}
\end{center}
\end{figure}

Diffusion experiments were performed using fluorescent correlation spectroscopy (see Supporting Information) with samples containing 10 nM labeled aldolase in the presence of varied concentrations of fructose-1,6-bisphosphate (FBP, 0-1 mM). In the absence of substrate, the diffusion coefficient of aldolase molecules was $D_0 = 42.6 \pm 1.0~\mu\text{m}^2\cdot \text{s}^{-1}$. We show in Fig. \ref{fig_exp_substrate}a the diffusion coefficient $D$ as a function of the concentration of substrate. The diffusion coefficient of the aldolase molecules was found to increase in a substrate concentration dependent manner, with relative enhancement that can reach up to 30\%. In order to rule out the possibility of de-agglomeration causing the enhanced diffusion of aldolase, we also compared the diffusion of aldolase before, during, and at the completion of the reaction. As shown in Fig. \ref{fig_exp_substrate}b, while the diffusivity of aldolase increases during turnover, it returns to the base value after the substrate is consumed.

The observed enhanced diffusion of aldolase with similar relative magnitudes to the significantly faster enzymes and the same characteristic Michaelis-Menten dependence on the substrate concentration poses an apparent paradox: the enhanced diffusion cannot be controlled by the magnitude of the reaction rate but it exhibits the same dependence on the substrate concentration. Moreover, given the thermodynamic properties of aldolase, the non-equilibrium mechanisms relying on the exothermicity of the catalytic reaction cannot be extended to the present case. Therefore, our experimental observations lead us to reconsider the theoretical paradigm around this physical phenomenon. First, it is necessary to determine if this enhancement is due to an intrinsically non-equilibrium process, or, in other words, if it is proportional to (or at least controlled by) the overall rate of catalysis. Secondly, we need to identify a  mechanism that would provide quantitative answers to account for the observed order of magnitude for the diffusion enhancement.

\begin{figure}
\begin{center}
\includegraphics[width=\columnwidth]{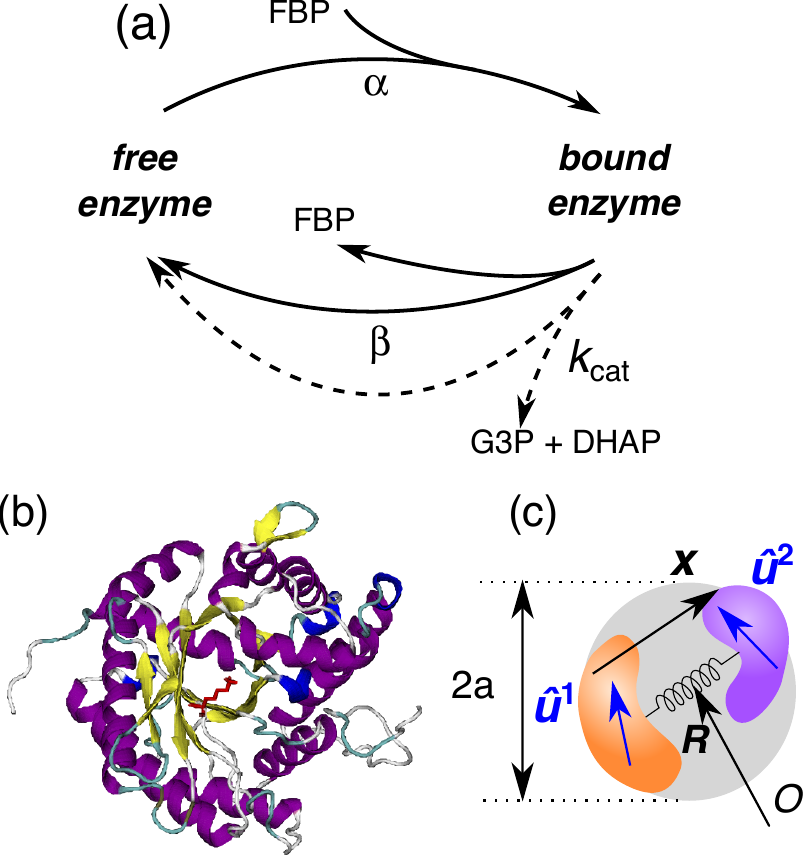}
\caption{(a) Substrate binding and unbinding drives a stochastic two-state process. The enzyme switches randomly between two equilibrium states where it is either free or bound.  (b) Structure of an aldolase monomer (Protein Data Bank ID: 1ADO, subunit A), generated with VMD \cite{Humphrey1996}. The residue colored in red indicates the location of the active site \cite{Rago2015}. (c) Aldolase enzyme modeled as a dumbbell. $\RR$ is the position of the centre of mass of the enzyme, $\xx$ represents its elongation. The grey sphere symbolizes the whole enzyme, whose typical size is denoted by $a$.}
\label{fig_theory}
\end{center}
\end{figure}

The first step in our modeling consists in a careful analysis of the relevant timescales of the phenomenon. Our approach is motivated by recent studies of enzyme conformational changes  \cite{Henzler-Wildman2007,Henzler-Wildman2007a}, and in particular, aldolase reaction pathways using fluorescence emission spectrophotometry \cite{Rago2015}, which have revealed that the rates of conformational changes could be much higher than the actual chemical rate, and reach values up to 10 to 100 s$^{-1}$.  It is important to take account of how the many competing time scales exist in the problem. The time scale for the actual conformational changes when they are triggered is of the order of the rotational diffusion time of the protein and is the shortest time scale in the system. The time scales for binding and unbinding of the substrate, which are purely physical processes at equilibrium since they do not involve subsequent conversion into products, are longer than the time scale for conformational changes but still shorter than the time scale for chemical conversion. Since the overall catalytic reaction is much slower than the conformational fluctuations, it is reasonable to neglect the chemical step of the cycle altogether. Consequently, we assume that the protein exists in two different states, namely a free state and a bound state, in which a substrate molecule is present in the active site (see Fig. \ref{fig_theory}a). Note that this simplified picture is an equilibrium description of the problem,  which does not involve the chemical or catalytic step of the process, and is therefore independent of the degree of exothermicity of the overall reaction.

\begin{figure}
\begin{center}
\includegraphics[width=0.65\columnwidth]{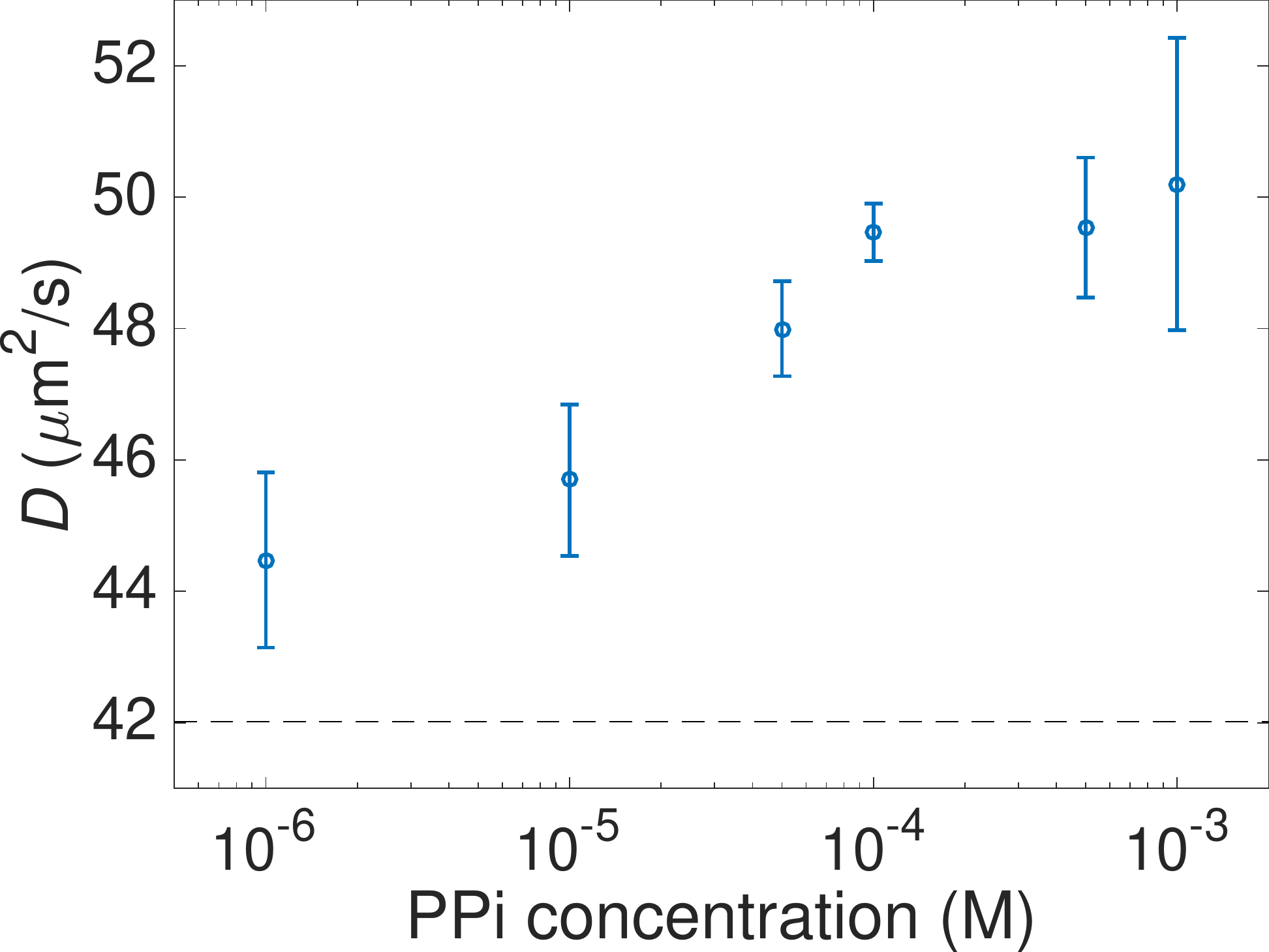}
\caption{Diffusion of aldolase enhances with increasing pyrophosphate (PPi)  concentration (the dashed line corresponds to the base value in the absence of substrate). PPi is a competitive inhibitor of aldolase.}
\label{inhibitor}
\end{center}
\end{figure}

In order to probe the importance of the catalytic step of the mechanochemical cycle, we have also measured diffusion of aldolase in the presence of pyrophosphate (PPi), which is a competitive inhibitor of aldolase and binds at the same active sites as FBP \cite{Rose1969,Callens1991}.  In the presence of PPi alone, diffusion of aldolase shows significant enhancement (Fig. \ref{inhibitor}), demonstrating that the catalytic step of the reaction scheme is not necessary to lead to enhanced diffusion. These findings are consistent with recent experiments performed on citrate synthase and malate dihydrogenase, which suggest that the diffusion coefficients of the enzymes are enhanced in the presence of their substrates even in the absence of their cofactors \cite{Wu2015}, and that binding/unbinding is sufficient to lead to enhanced diffusion of enzymes.

Relying on this simplified stochastic picture, we then aim to describe the effect of changes in conformational fluctuations induced by the binding and unbinding events. We first consider the simple case where the enzyme is always free (in the absence of substrate molecules). The state of the enzyme is then completely described by the position of its centre of mass $\RR$ and a vector  $\CC$, that describes the conformation of the enzyme, and whose dimension corresponds to the number of internal degrees of freedom. Given the complexity of the real structure of biomolecules (see Fig. \ref{fig_theory}b for a representation of aldolase), $\CC$ is a high-dimensional vector that does not need to be specified for now. The mobility coefficient $\mu$ of the enzyme depends on its geometrical properties, and therefore on its conformational state $\CC$. The overall diffusion coefficient of the enzyme as measured in the FCS experiments is an average on the conformations explored by the enzyme and can be related to the mobility through the fluctuation-dissipation theorem \cite{Einstein1905a} as
\begin{equation}
\label{ }
D= \kB T \int_{\CC} \mu(\CC)p(\CC) \equiv \kB T \moy{\mu},
\end{equation}
where $p(\CC)$ is the probability to find the enzyme in a given conformation $\CC$.

In the presence of the substrate, the enzyme switches randomly between a free state and another state where it is bound to a substrate molecule. The binding rate $\alpha$ is expected to be proportional to the substrate concentration $S$. Noticing that the distribution of the conformation coordinate $\CC$ is different in the two states (free and bound), we expect the rates $\alpha$ and $\beta$ to depend on this coordinate. The detailed balance condition takes the form $\alpha \,  \rho_\text{f}(\CC) = \beta \,  \rho_\text{b}(\CC)$, where $\rho_\text{f}$ (resp. $\rho_\text{b}$) is the distribution of $\CC$ knowing that the enzyme is in the free (resp. bound) state. Writing $\rho_\text{f}\propto \ex{-U_\text{f}/\kB T}$ and $\rho_\text{b}\propto \ex{-U_\text{b}/\kB T}$, where $U_\text{f}$ and $U_\text{b}$ are the effective potential corresponding to given conformations, we get
\begin{equation}
\label{ }
\frac{\alpha}{\beta} \propto \frac{S}{K_0} \ex{-[U_\text{b}(\CC)-U_\text{f}(\CC)]/\kB T},
\end{equation}
where $K_0$ is the bare equilibrium constant. The transitions of the enzyme between two equilibrium states therefore modify the effective distribution of the conformational variable. Assuming that the binding and unbinding rates $\alpha$ and $\beta$ are very large compared to the intrinsic timescales of the enzyme, one can establish the effective distribution of $\CC$ as
\begin{equation}
\label{ }
p(\CC) \simeq \frac{1}{Z} \left[ 1+\frac{S}{K_0} \ex{-[U_\text{b}(\CC)-U_\text{f}(\CC)]/\kB T} \right]\ex{-U_\text{f}(\CC)]/\kB T},
\end{equation}
where $Z$ is a normalization constant. It follows that the average of any conformation-dependent quantity $\Phi(\CC)$ can be written as
\begin{equation}
\label{ }
\moy{\Phi} = \moy{\Phi}_\text{f}+\left[\moy{\Phi}_\text{b}-\moy{\Phi}_\text{f} \right]\frac{S}{S+K},
\end{equation}
where the averages $\moy{\Phi}_\text{f}$ and $\moy{\Phi}_\text{b}$ are defined using the corresponding Boltzmann weights  $\ex{-U_\text{f}/\kB T}$ and $\ex{-U_\text{b}/\kB T}$, and where we define the equilibrium constant $K = K_0 \frac{\int_{\CC}\ex{-U_\text{f}(\CC)/\kB T}}{\int_{\CC}\ex{-U_\text{b}(\CC)/\kB T}}$. Within this picture, the relative diffusion enhancement writes
\begin{equation}
\label{deltaD}
\frac{\Delta D}{D_0} = \frac{\moy{\mu}_\text{b}-\moy{\mu}_\text{f}}{\moy{\mu}_\text{f}}  \frac{S}{S+K} \equiv \mathcal{A}\,  \frac{S}{S+K} .
\end{equation}
This result shows that even if the catalytic step of the chemical cycle is neglected, in such a way that the modifications of the diffusion coefficient cannot be related to the rate of product formation, the relative change in  diffusion  still exhibits a Michaelis-Menten-like dependence over the substrate concentration, and is independent of the catalytic rate of the whole chemical reaction. The dimensionless coefficient $\mathcal{A}$ is  a complex quantity, that depends on the shape of the interaction potentials $U_\text{f}(\CC)$ and $U_\text{b}(\CC)$, and that includes contributions from all the internal degrees of freedom of the enzyme that are affected by binding and unbinding. 
\begin{figure}
\begin{center}
\includegraphics[width=\columnwidth]{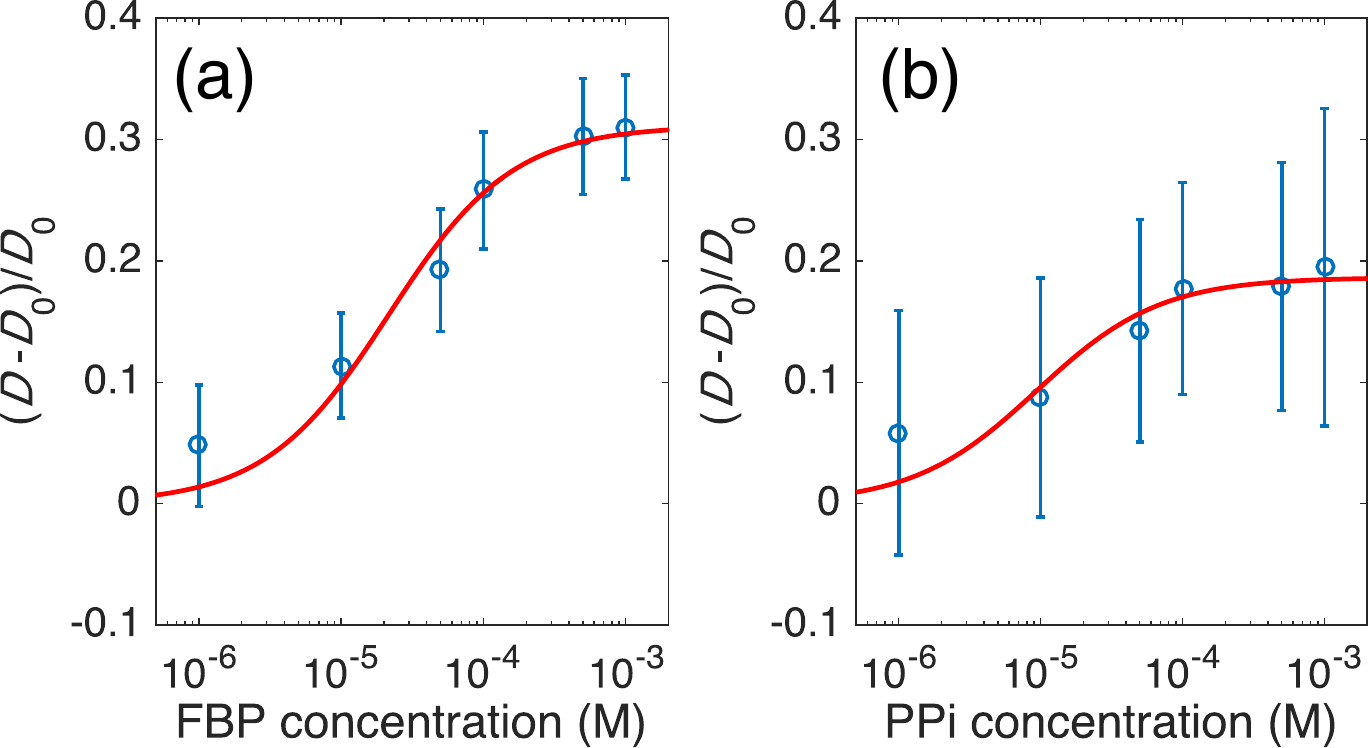}
\caption{Relative increase of the diffusion coefficient of aldolase molecules measured in FCS experiments (symbols) in the presence of: (a) FBP as a substrate, (b) pyrophosphate (PPi) as a competitive inhibitor,  and compared to the fitting function $\frac{\Delta D}{D_0} = \mathcal{A} \frac{S}{K+S}$ (solid line), with $\mathcal{A}$ and $K$ as free parameters. The relatively large error bars for the experiments performed with PPi originate from the error on the measurement of $D_0$, which affects the standard deviation for the quantity $\Delta D/D_0$.}
\label{fig_fit}
\end{center}
\end{figure}
This simple equation \eqref{deltaD}, which contains the minimal ingredients of our new physical paradigm, can be used to fit the experimental data obtained for aldolase in the presence of the substrate FBP or in the presence of the competitive inhibitor PPi with $\mathcal{A}$ and $K$ as free parameters. For the experiments with FBP (Fig. \ref{fig_fit}a), we find $\mathcal{A}=0.3$ and $K= 2.16\cdot10^{-5}$ M, which is comparable to the Michaelis constant reported for aldolase in the presence of FBP at physiological pH \cite{Callens1991} ($1.28\cdot10^{-5}$ M). In the presence of the inhibitor PPi (Fig. \ref{fig_fit}b), we find $\mathcal{A}=0.19$ and $K= 9.4\cdot 10^{-6} $ M, the dissociation constant given in the literature being $4.6\cdot 10^{-5}$ M \cite{Kasprzak1980}.

In order to get a more quantitative description of the changes in the averages mobility coefficients $\moy{\mu}_\text{f}$ and $\moy{\mu}_\text{b}$, we need to consider in greater details the modifications of the conformational fluctuations induced by binding and unbinding. The simplest way to describe the conformational state of the enzyme is to reduce the conformational state $\CC$ to a single parameter $R$ that describes the hydrodynamic radius of the enzyme. Structural studies of aldolase have recently shown that the effect of FBP binding was to bring residues near the active site closer one to another \cite{Rago2015}, therefore effectively reducing the hydrodynamic radius of the molecule. The mobility coefficient of the molecule goes as the inverse of the hydrodynamic radius, so that the contribution to $\mathcal{A}$ coming from this effective size reduction can be estimated as $\mathcal{A}_1 \sim |\delta R|/R$, in a way that deformations of the order of a few \AA~can have a significant impact on the measured diffusion coefficient. However, this effect cannot account on its own for the order of magnitude of the diffusion enhancement.

Then, in order to go further in the description of the internal degrees of freedom of the molecule and to take into account the effect of binding/unbinding on its elastic properties, we use a minimal dumbbell model (Fig. \ref{fig_theory}c), where the structure of the enzyme is reduced to two hydrodynamically coupled subunits interacting via a harmonic potential, and that we recently studied in detail \cite{Illien2016}. The conformational state $\CC$ now reduces to a vector $\xx$ that represents the elongation of the dumbbell. In the particular case where the dumbbell is symmetric, and in the limit where the subunits are far one from another, one can show that the averaged mobility coefficient is given by \cite{Illien2016}
\begin{equation}
\label{ }
\moy{\mu} = \frac{1}{6\pi\eta} \left( \frac{1}{2 a_0} + \frac{3}{8}\moy{\frac{1}{x}}  \right) ,
\end{equation}
where $\eta$ is the viscosity of water, $a_0$ the typical size of the subunits that  $x$ is the length of the dumbbell. Binding of a substrate molecule to the enzyme will generally hinder the fluctuations of internal degrees of freedom, and therefore  make the protein stiffer. The contribution $\mathcal{A}_2$ to the dimensionless coefficient $ \left[ \moy{\mu}_\text{b}-\moy{\mu}_\text{f} \right] / \moy{\mu}_\text{f} $ can be calculated explicitly by assuming that the potential energies associated to the internal variable $x$ are of the form $U_\text{f}=\frac{1}{2}k_\text{f}(x-a)^2$ and $U_\text{b}=\frac{1}{2}k_\text{b}(x-a)^2$, with $k_\text{b}>k_\text{f}$ and where $a$ is the typical size of the enzyme. In the limit of very large $k_\text{f}$ and $k_\text{b}$ with a finite difference $\delta k \equiv k_\text{b}-k_\text{f} $, we find $\mathcal{A}_2 \propto \frac{\kB T}{k_\text{f} a^2} \frac{\delta k}{k_\text{f}}$ up to a dimensionless prefactor of order 1. The dimensionless number $\frac{\kB T}{k_\text{f} a^2}$ represents the relative amplitude of the length fluctuations of the dumbbell, and is bounded by unity, such that increased stiffness can significantly increase the enzyme diffusion coefficient. This contribution can be related to the concept of entropic allostery \cite{Hawkins2004}, which suggests that ligand binding to a macromolecule can change its vibrational entropy, in addition to affecting its static structure.

Finally, this model can be refined by assuming that the subunits have more complex shapes and undergo orientational fluctuations (Fig. \ref{fig_theory}c). The conformation state of the enzyme is then described by the vector $\CC=(\xx,\uu^1,\uu^2)$ where $\uu^1$ and $\uu^2$ are unit vectors characterising the orientations of the subunits. These additional degrees of freedom will affect the overall diffusion coefficient of the dumbbell. We recently employed a Fokker-Planck description of the stochastic dynamics of the dumbbell and a careful treatment of the coupling between the internal and external degrees of freedom induced by hydrodynamic interactions to show that the internal fluctuations contribute negatively to the effective diffusion coefficient of the position of the dumbbell \cite{Illien2016}. It is beyond the scope of this paper to present the details of this calculation, and we simply give the following simplified and generic form for the effective diffusion coefficient:
\begin{equation}
\label{ }
D = D_\text{ave}-\delta D_\text{fluc},
\end{equation}
where  the first term corresponds to the  average contribution from the translational modes of the dumbbell, and the second term represents fluctuation--induced corrections arising from the internal degrees of freedom. The latter is controlled by the asymmetry of the dumbbell and the anisotropy of the individual subunits, and is typically a fraction of $D_\text{ave}$, depending on the precise geometrical properties of the dumbbell. Its negative sign is a generic feature of fluctuation--induced interactions \cite{Kardar1999}.  In particular, this analysis indicates that hindering the orientational fluctuations of freely-rotating parts of the molecule can  enhance its overall diffusion. A more detailed theoretical study of this effect will be the object of a later publication.

Therefore, these contributions, that originate from a reduction of the hydrodynamic radius of the enzyme, an increased stiffness, or hindering of the internal modes fluctuations of the enzyme can yield significant diffusion enhancements, which are of the order of a fraction of the bare diffusion coefficient of the enzyme. Although this extended dumbbell model is an idealized representation of the enzyme that greatly simplifies its structure, it contains, with very few internal degrees of freedom, the minimal ingredients to represent the compressional and orientational fluctuation modes that prevail inside a real macromolecule, and should therefore accurately predict the main features observed with FCS experiments.

In summary, by employing aldolase, a slow enzyme that catalyzes an endothermic reaction, we demonstrated experimentally that exothermicity is not a necessary condition for the observation of enhanced diffusion in the presence of substrate molecules. These results challenge previous physicals scenarios that were proposed to account for this phenomenon, and that only held when the amount of heat released by the enzyme at each catalytic turnover was significant, or when the overall catalytic rate was sufficiently large. Guided by these experimental results and by structural studies of aldolase, we proposed a new physical paradigm, in which the enzyme stochastically switches between two equilibrium states, in which it is either free or bond. Considering that binding and unbinding significantly affects the conformational fluctuations of the enzyme, we were able to measure the change in its diffusion coefficient as measured in FCS experiments in terms of its the averaged mobility coefficients. Using simple physical arguments and a more subtle analysis of the fluctuation--induced effects mediated by hydrodynamic interactions, we generically show how substrate binding can modify the mobility and eventually enhance the diffusion of the enzyme.

While we have obtained this result using the assumption that the binding and unbinding rates are considerably higher than the catalytic reaction rate, it is natural to expect that for faster enzymes these rates could be comparable, in which case we will obtain a combination of the above effect and the stochastic swimming that is controlled by the (fast) reaction rate.  This picture, inspired by a biological system, constitutes a new physical phenomenon, that was overlooked so far. Finally, we emphasize the generality of this mechanism; since substrate binding-unbinding is universal for enzymes, the proposed mechanism for enhanced diffusion should be universally present for all enzymes, and should be observable provided the changes in the conformational fluctuations are sufficiently large in relative terms. While our main aim has been to propose a new generic physical mechanism, more detailed studies of the molecular structure of the enzymes, for example using molecular dynamics simulation \cite{Echeverria2011}, could help determine the precise characteristics that would allow enhanced diffusion of enzymes upon substrate binding and unbinding.

%\bibliographystyle{achemso}
%\bibliography{/Users/pierreillien/Documents/library.bib}

\providecommand{\latin}[1]{#1}
\providecommand*\mcitethebibliography{\thebibliography}
\csname @ifundefined\endcsname{endmcitethebibliography}
  {\let\endmcitethebibliography\endthebibliography}{}

\onecolumngrid

\pagebreak

\pagebreak

\begin{center}

{\large \textbf{Exothermicity is not a necessary condition for enhanced diffusion of enzymes}}

$\ $

{\large \textbf{\textit{Supporting Information}}}

$\ $

Pierre Illien,$^{1,2}$\footnote[1]{P.I. and X.Z. contributed equally to this work} Xi Zhao,$^2$$^*$ Krishna K. Dey,$^2$\footnote[2]{Present address: Indian Institute of Technology Gandhinagar, Palaj Campus, Gandhinagar, Gujarat 382 355, India.} Peter J. Butler,$^3$ Ayusman Sen$^2$\footnote[3]{Corresponding authors: asen@psu.edu, ramin.golestanian@physics.ox.ac.uk} and Ramin Golestanian$^1$$^\ddag$

$ \ $

$^1$\textit{Rudolf Peierls Centre for Theoretical Physics, University of Oxford, Oxford OX1 3NP, UK}

$^2$\textit{Department of Chemistry, The Pennsylvania State University, University Park, PA 16802, USA}

$^3$\textit{Department of Biomedical Engineering, The Pennsylvania State University, University Park, PA 16802, USA}

\end{center}

%\section{Experimental methods}

\section{Tagging Enzymes with Dylight 550 Maleimide }

Fructose bisphosphate aldolase (from rabbit muscle; Sigma-Aldrich) was labeled with a thiol-reactive dye Dylight 550 (ex/em: 557/572; Thermo Fisher Scientific). Labeling of aldolase (7.5 $\mu$M) was carried out with four fold excess of the fluorescent dye (31.2 $\mu$M) and 1.27 mM EDTA in an ice bath for 3 h in 50 mM Hepes buffer (pH 7.4). The enzyme-dye complexes were further purified using Amicon Ultra-4 Centrifugal Filter Unit with Ultracel-10 membrane (EMD Millipore) with 50 mM HEPES buffer (pH 7.4) to minimize the free-dye concentration in solution. The tagged enzyme concentration and degree of labeling were calculated spectrophotometrically, with a typical dye/protein ratio of 1:1.

\section{Diffusion Measurement using FCS}

%Autocorrelation curves were fitted to Eq. \ref{FCSfit} using Levenberg-Marquardt non-linear least squares regression algorithm with Origin software. Quality of the fitted curves was assessed based on chi-square ($\chi^2$) values. 

%Diffusion coefficients of the fluorescent enzymes were measured in the presence of enzymatic catalysis using TCSPC instrumentation. Fluctuations in fluorescence intensity arising from the diffusion of the tracer particles were auto-correlated and fitted to a multi-component 3D diffusion model to determine their diffusion coefficient in solution. The calibration solution was prepared in DI water. The experimental enzyme solutions were prepared with HEPES buffer. Using the oscilloscope in the SPC-630 module from Becker and Hickl, the laser was focused to be within the solution, where multiple measurements were taken using the FIFO mode.

Diffusion coefficients of the fluorescent enzymes were measured in the presence of enzymatic catalysis using TCSPC instrumentation. Fluctuations in fluorescence intensity arising from the diffusion of fluorescently labeled enzymes were auto-correlated and fitted to a multi-component 3D diffusion model to determine their diffusion coefficient in solution. The experimental enzyme solutions were prepared with HEPES buffer. Using the oscilloscope in the SPC-630 module from Becker and Hickl, the laser was focused to be within the solution, where multiple measurements were taken using the FIFO mode.

FCS was performed on a custom-built microscope based optical setup. Excitation light from a PicoTRAIN 532 nm, 80 MHz, 5.4 ps pulsed laser (High-Q Laser) was guided through an IX-71 microscope (Olympus), with an Olympus 60$\times$/1.2-NA water-immersion objective. Emitted fluorescent light from the sample was passed through a dichroic beam splitter (Z520RDC-SP-POL, Chroma Technology) and focused onto a 50 $\mu$m, 0.22-NA optical fibre (Thorlabs), which acted as a confocal pinhole. The signal from the photomultiplier tube was routed to a preamplifier (HFAC-26) and then to a time-correlated single-photon counting (TCSPC) board (SPC-630, Becker and Hickl). The sample was positioned with a high-resolution 3-D piezoelectric stage (NanoView, Mad City Laboratories). The measurements were performed with 26.5 $\mu$W and 30.2 $\mu$W excitation power for tagged enzyme and tracers, respectively, and the optical system was calibrated before each experiment using free 50 nm fluorescent particles in double distilled water. Fluorescent molecules moving in and out of the diffraction-limited observation volume induce bursts in fluorescence collected in first-in, first-out (FIFO) mode by the TCSPC board, which was incorporated in the instrument. Fluctuations in fluorescence intensity from the diffusion of molecules were auto-correlated and fit by a multicomponent 3D model to determine the diffusion coefficients of individual species. The autocorrelation of the intensity signal is given by \cite{Krichevsky2002}
\begin{equation}
\label{FCSfit}
G(\tau) =  \frac{1}{N}\left[ 1+ \frac{\tau}{\tau_D}      \right]^{-1} \left[ 1+\frac{1}{w^2} \frac{\tau}{\tau_D}  \right]^{-1/2}
\end{equation}
where $\tau_D = r^2/(4 D)$. Here, $N$ is the average number of fluorophores in the observation volume, $\tau$ is the auto-correlation time, $w$ is the structure factor, which is defined as the ratio of height to width of the illumination profile, and $\tau_D$ is the characteristic diffusion time of the fluorescent particle with diffusion coefficient $D$ crossing a circular area with radius $r$. 

The quality of the fitted curves was assessed by chi-square ($\chi^2$) analysis. Typical values of $w$ and $r$ obtained during one calibration (comprising of 5 independent measurements, carried out under identical conditions) is shown in Table \ref{table1}.
 
  \begin{table}
 \begin{center}
  \begin{tabular}{|c|c|c|c|c|c|}
  \hline
 Measurement	 & $\tau_D$ ($\mu$s)  &     $N$  &	$r$ (nm)	&  w  &	$\chi^2$    \\		\hline
1	&	7138.25	&	0.89	&	505.80	&	7.50	&	0.00018	\\		\hline
2	&	7648.94	&	0.85	&	523.58	&	8.35	&	0.00022	\\		\hline
3	&	7741.99	&	0.91	&	526.76	&	10.42	&	0.00028	\\		\hline
4	&	6676.06	&	1.05	&	489.15	&	6.03	&	0.00005	\\		\hline
5	&	7244.59	&	1.11	&	509.55	&	5.65	&	0.00006	\\		\hline
Average	&	7289.97	&	0.96	&	510.97	&	7.59	&	0.00016	\\		\hline
St. Dev.	&	428.70	&	0.11	&	15.11	&	1.92	&	0.00010	\\		\hline
\end{tabular}
\caption{Typical  calibration parameters obtained in FCS experiments}
\label{table1}
  \end{center}
  \end{table}

Next, fluctuations in fluorescence intensity arising from the diffusion of fluorescently tagged enzyme molecules within the confocal volume in the presence and absence of substrate FBP, were auto-correlated to Eq. \ref{FCSfit} to determine the characteristic diffusion time, and thereby the translational diffusion constant $D$ of the molecule, using the values of $w$ and $r$ measured during calibration.
 
\begin{figure}
\begin{center}
\includegraphics[width=8cm]{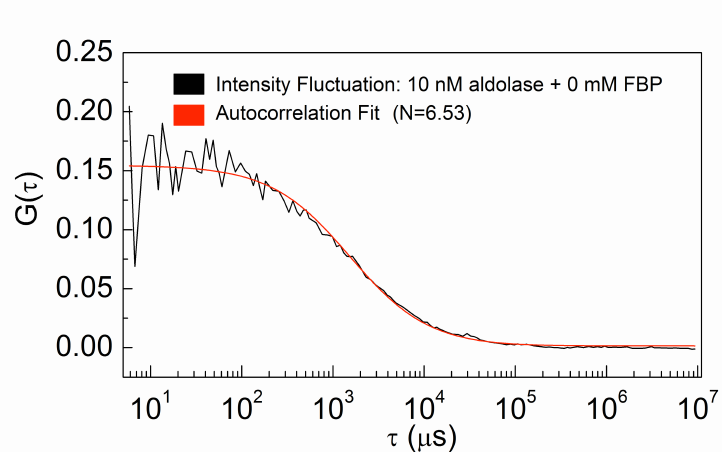}
\includegraphics[width=8cm]{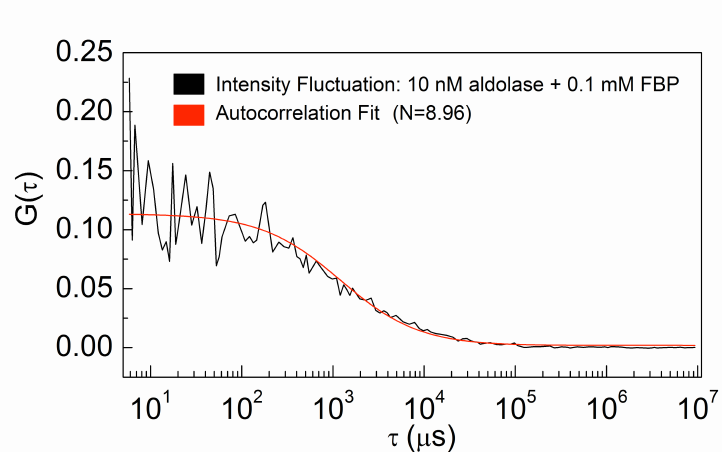}
\caption{Representative intensity fluctuation curves and corresponding autocorrelation fits obtained for fluorescently tagged aldolase dispersed in 0 mM and 0.1 mM FBP solutions respectively}
\label{FCSplot}
\end{center}
\end{figure}

The quality of the fitted curves was again assessed by chi-square ($\chi^2$) analysis. Figure \ref{FCSplot} shows representative intensity fluctuation curves and corresponding autocorrelation fits obtained for fluorescently tagged aldolase dispersed in 0 mM and 0.1 mM FBP solutions respectively. 
 
In Table \ref{table2}, we show typical values of $\tau_D$ obtained for tagged aldolase dispersed in different concentrations of FBP. Each experiment comprises of 5 independent measurements, carried out under identical conditions.  
 
\begin{table}
 
\begin{tabular}{ccc}
 
\begin{tabular}{|c|c|c|c|}
\multicolumn{4}{c}{\textbf{[FBP] = 0 mM}}\\
  \hline
 Measurement	 & $\tau_D$ ($\mu$s)  &     $N$ &	$\chi^2$    \\		\hline
1	&	1574.37	&	6.38	&	0.00019\\		\hline
2	&	1608.66	&	5.38	&	0.00018\\		\hline
3	&	1529.10	&	6.53	&	0.00014\\		\hline
4	&	1552.88	&	5.18	&	0.00043\\		\hline
5	&	1547.65	&	7.49	&	0.00016\\		\hline
Average	&	1562.53	&	6.19	&	0.00022\\		\hline
St. Dev.	&	30.41	&	0.94	&	0.00012\\		\hline
\end{tabular}

&

\begin{tabular}{|c|c|c|c|}
\multicolumn{4}{c}{\textbf{[FBP] = 0.001 mM}}\\
  \hline
 Measurement	 & $\tau_D$ ($\mu$s)  &     $N$ &	$\chi^2$    \\		\hline
1	&	1569.76	&	7.30	&	0.00014 \\		\hline
2	&	1440.32	&	5.63	&	0.00017 \\		\hline
3	&	1443.54	&	6.12	&	0.00025 \\		\hline
4	&	1480.51	&	6.10	&	0.00037 \\		\hline
5	&	1489.36	&	7.06	&	0.00041 \\		\hline
Average	&	1484.70	&	6.44	&	0.00027 \\		\hline
St. Dev.	&	52.29	&	0.71	&	0.00012 \\		\hline
\end{tabular}

&

  \begin{tabular}{|c|c|c|c|}
\multicolumn{4}{c}{\textbf{[FBP] = 0.01 mM}}\\
  \hline
 Measurement	 & $\tau_D$ ($\mu$s)  &     $N$ &	$\chi^2$    \\		\hline
1	&	1364.61	&	5.83	&	0.00027\\		\hline
2	&	1348.07	&	5.19	&	0.00058\\		\hline
3	&	1431.66	&	4.60	&	0.00022\\		\hline
4	&	1379.26	&	5.50	&	0.00037\\		\hline
5	&	1366.80	&	5.09	&	0.00041\\		\hline
Average	&	1378.08	&	5.24	&	0.00037\\		\hline
St. Dev.	&	31.94	&	0.46	&	0.00014\\		\hline
\end{tabular}

\end{tabular}

\begin{tabular}{ccc}

  \begin{tabular}{|c|c|c|c|}
\multicolumn{4}{c}{\textbf{[FBP] = 0.05 mM}}\\
  \hline
 Measurement	 & $\tau_D$ ($\mu$s)  &     $N$ &	$\chi^2$    \\		\hline
1	&	1304.24		&	7.19		&	0.00019 \\		\hline
2	&	1297.88		&	6.86		&	0.00025 \\		\hline
3	&	1236.82		&	8.13		&	0.00017 \\		\hline
4	&	1303.69		&	7.59		&	0.00023 \\		\hline
5	&	1299.18		&	11.5		&	0.00022 \\		\hline
Average 	&	1288.36		&	8.25		&	0.00021 \\		\hline
St. Dev. 	&	28.95		&	1.88		&	0.00003 \\		\hline
\end{tabular}
 
&
 
\begin{tabular}{|c|c|c|c|}
\multicolumn{4}{c}{\textbf{[FBP] = 0.1 mM}}\\
\hline
 Measurement	 & $\tau_D$ ($\mu$s)  &     $N$ &	$\chi^2$    \\		\hline
1	&	1219.51	&	6.72	&	0.00014    \\		\hline
2	&	1278.62	&	7.66	&	0.00026    \\		\hline
3	&	1202.28	&	7.97	&	0.00032    \\		\hline
4	&	1225.02	&	9.00	&	0.00025    \\		\hline
5	&	1205.04	&	8.96	&	0.00029    \\		\hline
Average	&	1226.09	&	8.06	&	0.00025    \\		\hline
St. Dev.	&	30.88	&	0.96	&	0.00007    \\		\hline
\end{tabular}
 
 &
  \begin{tabular}{|c|c|c|c|}
\multicolumn{4}{c}{\textbf{[FBP] = 0.5 mM}}\\
  \hline
 Measurement	 & $\tau_D$ ($\mu$s)  &     $N$ &	$\chi^2$    \\		\hline
1	&	1167.97		&	8.10		&	0.00016    \\		\hline
2	&	1200.14		&	4.89		&	0.00047    \\		\hline
3	&	1166.14		&	4.21		&	0.00040    \\		\hline
4	&	1186.99		&	6.21		&	0.00027    \\		\hline
5	&	1194.38		&	5.30		&	0.00044    \\		\hline
Average 	&	1183.12		&	5.74		&	0.00035    \\		\hline
St. Dev. 	&	15.41		&	1.50		&	0.00013    \\		\hline

\end{tabular}
 
\end{tabular}

\begin{tabular}{|c|c|c|c|}
\multicolumn{4}{c}{\textbf{[FBP] = 1 mM}}\\
  \hline
 Measurement	 & $\tau_D$ ($\mu$s)  &     $N$ &	$\chi^2$    \\		\hline
1	&	1175.07	&	6.70	&	0.00022    \\		\hline
2	&	1181.49	&	4.34	&	0.00041    \\		\hline
3	&	1164.99	&	4.76	&	0.00090    \\		\hline
4	&	1176.45	&	3.93	&	0.00077    \\		\hline
5	&	1170.87	&	4.75	&	0.00034    \\		\hline
Average	&	1173.74	&	4.90	&	0.00053    \\		\hline
St. Dev.	&	6.20	&	1.06	&	0.00029    \\		\hline
\end{tabular}

\caption{Typical  parameters obtained in FCS experiments with different FBP concentrations}
\label{table2}
\end{table}

\providecommand{\latin}[1]{#1}
\providecommand*\mcitethebibliography{\thebibliography}
\csname @ifundefined\endcsname{endmcitethebibliography}
  {\let\endmcitethebibliography\endthebibliography}{}

\end{document}